%-----------------------------------------------------------------------
\overfullrule=0pt
%-----------------------------------------------------------------------
%Nature referencing macro
%-----------------------------------------------------------------------
\def\refto#1{$^{#1}$}           % For references in text as superscript
\def\ref#1{ref.~#1}                     %       for inline references
\def\Ref#1{#1}                          %       ditto
\gdef\refis#1{\item{#1.\ }}                    % Ref list numbers.
\def\beginparmode{\endmode
  \begingroup \def\endmode{\par\endgroup}}
\let\endmode=\par
\def\body{\beginparmode}
\def\head#1{                    % Head;  NOTE enclose the text in {}
  \goodbreak\vskip 0.5truein    %  e.g., \head{I. Introduction}
  {\centerline{\bf{#1}}\par}
   \nobreak\vskip 0.25truein\nobreak}
\def\references                 % Begin references -- basic format is
Phys Rev
  {\head{References}            % I.e., volume, page, year (space after commas)\.
   \beginparmode
   \frenchspacing \parindent=0pt \leftskip=1truecm
   \parskip=8pt plus 3pt \everypar{\hangindent=\parindent}}
\def\endreferences{\body}
%-----------------------------------------------------------------------
%rji's referencing macro
%-----------------------------------------------------------------------
\def\gs{\mathrel{\raise0.35ex\hbox{$\scriptstyle >$}\kern-0.6em
\lower0.40ex\hbox{{$\scriptstyle \sim$}}}}
\def\ls{\mathrel{\raise0.35ex\hbox{$\scriptstyle <$}\kern-0.6em
\lower0.40ex\hbox{{$\scriptstyle \sim$}}}}
\def\kms{km$\,$s$^{-1}$}

\def\sec{{$\,$s}}

%------------------------------------------------------------------------
%Other needed macro files  (Site dependent)
%------------------------------------------------------------------------
%\input reforder
%\input citmac
\catcode`@=11
\newcount\r@fcount \r@fcount=0
\newcount\r@fcurr
\immediate\newwrite\reffile
\newif\ifr@ffile\r@ffilefalse
\def\w@rnwrite#1{\ifr@ffile\immediate\write\reffile{#1}\fi\message{#1}}

\def\writer@f#1>>{}
\def\referencefile{%			  Stuff to write .REF file
  \r@ffiletrue\immediate\openout\reffile=\jobname.ref%
  \def\writer@f##1>>{\ifr@ffile\immediate\write\reffile%
    {\noexpand\refis{##1} = \csname r@fnum##1\endcsname = %
     \expandafter\expandafter\expandafter\strip@t\expandafter%
     \meaning\csname r@ftext\csname r@fnum##1\endcsname\endcsname}\fi}%
  \def\strip@t##1>>{}}

\def\citeall#1{\xdef#1##1{#1{\noexpand\cite{##1}}}}
\def\cite#1{\each@rg\citer@nge{#1}}	% Variable No. of args, separated by ","

\def\each@rg#1#2{{\let\thecsname=#1\expandafter\first@rg#2,\end,}}
\def\first@rg#1,{\thecsname{#1}\apply@rg}	% each@ag is a general purpose
\def\apply@rg#1,{\ifx\end#1\let\next=\relax%	  variable no. of arg. macro.
\else,\thecsname{#1}\let\next=\apply@rg\fi\next}% args separated by commas

\def\citer@nge#1{\citedor@nge#1-\end-}	% Check for M-N range (M and N numbers)
\def\citer@ngeat#1\end-{#1}
\def\citedor@nge#1-#2-{\ifx\end#2\r@featspace#1 % Single argument
  \else\citel@@p{#1}{#2}\citer@ngeat\fi}	% M-N range of arguments
\def\citel@@p#1#2{\ifnum#1>#2{\errmessage{Reference range #1-#2\space is bad.}%
    \errhelp{If you cite a series of references by the notation M-N, then M and
    N must be integers, and N must be greater than or equal to M.}}\else%
 {\count0=#1\count1=#2\advance\count1 by1\relax\expandafter\r@fcite\the\count0,%
  \loop\advance\count0 by1\relax%	  Loop from M to N
    \ifnum\count0<\count1,\expandafter\r@fcite\the\count0,%
  \repeat}\fi}

\def\r@featspace#1#2 {\r@fcite#1#2,}	% Eat spaces at beginning or end of arg
\def\r@fcite#1,{\ifuncit@d{#1}%		  Cite individual reference
    \newr@f{#1}%
    \expandafter\gdef\csname r@ftext\number\r@fcount\endcsname%
                     {\message{Reference #1 to be supplied.}%
                      \writer@f#1>>#1 to be supplied.\par}%
 \fi%
 \csname r@fnum#1\endcsname}
\def\ifuncit@d#1{\expandafter\ifx\csname r@fnum#1\endcsname\relax}%
\def\newr@f#1{\global\advance\r@fcount by1%
    \expandafter\xdef\csname r@fnum#1\endcsname{\number\r@fcount}}

\let\r@fis=\refis			% Save old \refis, redefine
\def\refis#1#2#3\par{\ifuncit@d{#1}%      Use two params #2 #3 to strip blank
   \newr@f{#1}%
   \w@rnwrite{Reference #1=\number\r@fcount\space is not cited up to now.}\fi%
  \expandafter\gdef\csname r@ftext\csname r@fnum#1\endcsname\endcsname%
  {\writer@f#1>>#2#3\par}}

\def\ignoreuncited{%   redefine \refis if ignoring uncited references
   \def\refis##1##2##3\par{\ifuncit@d{##1}%
     \else\expandafter\gdef\csname r@ftext\csname r@fnum##1\endcsname\endcsname%
     {\writer@f##1>>##2##3\par}\fi}}

\def\r@ferr{\endreferences\errmessage{I was expecting to see
\noexpand\endreferences before now;  I have inserted it here.}}
\let\r@ferences=\references
\def\references{\r@ferences\def\endmode{\r@ferr\par\endgroup}}

\let\endr@ferences=\endreferences
\def\endreferences{\r@fcurr=0%		  Save old \endreferences, redefine
  {\loop\ifnum\r@fcurr<\r@fcount%	  Loop over refnum and produce text
    \advance\r@fcurr by 1\relax\expandafter\r@fis\expandafter{\number\r@fcurr}%
    \csname r@ftext\number\r@fcurr\endcsname%
  \repeat}\gdef\r@ferr{}\endr@ferences}

% Save old \endpaper, redefine it to write parting message.

\let\r@fend=\endpaper\gdef\endpaper{\ifr@ffile
\immediate\write16{Cross References written on []\jobname.REF.}\fi\r@fend}

\catcode`@=12

\citeall\refto		% These macros will generate citations
\citeall\ref		%
\citeall\Ref		%

%--------------------------------%
%  GENERALLY USEFUL DEFINITIONS  %
%--------------------------------%
\def\singlespace{\baselineskip 12pt \lineskip 1pt \parskip 2pt plus 1 pt}

\def\today{\number\day\enspace
     \ifcase\month\or January\or Febuary\or March\or April\or May\or
     June\or July\or August\or September\or October\or
     November\or December\fi \enspace\number\year}
\def\clock{\count0=\time \divide\count0 by 60
    \count1=\count0 \multiply\count1 by -60 \advance\count1 by \time
    \number\count0:\ifnum\count1<10{0\number\count1}\else\number\count1\fi}
\footline={\hss -- \folio\ -- \hss}

%-----------%
%  SYMBOLS  %
%-----------%
\def\deg{\ifmmode^\circ\else$^\circ$\fi}
\def\solar{\ifmmode_{\mathord\odot}\else$_{\mathord\odot}$\fi}
%--------------%
%  REFERENCES  %
%--------------%
\def\jref#1 #2 #3 #4 {{\par\noindent \hangindent=3em \hangafter=1 
      \advance \rightskip by 5em #1, {\it#2}, {\bf#3}, #4.\par}}
\def\ref#1{{\par\noindent \hangindent=3em \hangafter=1 
      \advance \rightskip by 5em #1.\par}}
%----------------------------------%
%  INCREMENTAL EQUATION NUMBERING  %
%----------------------------------%
\newcount\eqnum
\def\nexteq{\global\advance\eqnum by1 \eqno(\number\eqnum)}
\def\lasteq#1{\if)#1[\number\eqnum]\else(\number\eqnum)\fi#1}
\def\preveq#1#2{{\advance\eqnum by-#1
    \if)#2[\number\eqnum]\else(\number\eqnum)\fi}#2}
%----------%
%  TABLES  %
%----------%

\def\tableheight{\vrule width 0pt height 8.5pt depth 3.5pt}
{\catcode`|=\active \catcode`&=\active 
    \gdef\tabledelim{\catcode`|=\active \let|=\vbar
                     \catcode`&=\active \let&=\nobar} }
\def\table{\begingroup
    \def\twidth{\hsize}
    \def\tablewidth##1{\def\twidth{##1}}
    \def\defaultheight{\vrule width 0pt height 8.5pt depth 3.5pt}
    \def\heightdepth##1{\dimen0=##1
        \ifdim\dimen0>5pt 
            \divide\dimen0 by 2 \advance\dimen0 by 2.5pt
            \dimen1=\dimen0 \advance\dimen1 by -5pt
            \vrule width 0pt height \the\dimen0  depth \the\dimen1
        \else  \divide\dimen0 by 2
            \vrule width 0pt height \the\dimen0  depth \the\dimen0 \fi}
    \def\spacing##1{\def\defaultheight{\heightdepth{##1}}}
    \def\nextheight##1{\noalign{\gdef\tableheight{\heightdepth{##1}}}}
    \def\end{\cr\noalign{\gdef\tableheight{\defaultheight}}}
    \def\zerowidth##1{\omit\hidewidth ##1 \hidewidth}    
    \def\hline{\noalign{\hrule}}
    \def\skip##1{\noalign{\vskip##1}}
    \def\bskip##1{\noalign{\hbox to \twidth{\vrule height##1 depth 0pt \hfil
        \vrule height##1 depth 0pt}}}
    \def\header##1{\noalign{\hbox to \twidth{\hfil ##1 \unskip\hfil}}}
    \def\bheader##1{\noalign{\hbox to \twidth{\vrule\hfil ##1 
        \unskip\hfil\vrule}}}
    \def\spanloop{\span\omit \advance\mscount by -1}
    \def\extend##1##2{\omit
        \mscount=##1 \multiply\mscount by 2 \advance\mscount by -1
        \loop\ifnum\mscount>1 \spanloop\repeat \ \hfil ##2 \unskip\hfil}
    \def\vbar{&\vrule&}
    \def\nobar{&&}
    \def\hdash##1{ \noalign{ \relax \gdef\tableheight{\heightdepth{0pt}}
        \toks0={} \count0=1 \count1=0 \putout##1\end 
        \toks0=\expandafter{\the\toks0 &\end} \xdef\piggy{\the\toks0} }
        \piggy}
    \let\e=\expandafter
    \def\putspace{\ifnum\count0>1 \advance\count0 by -1
        \toks0=\e\e\e{\the\e\toks0\e&\e\multispan\e{\the\count0}\hfill} 
        \fi \count0=0 }
    \def\putrule{\ifnum\count1>0 \advance\count1 by 1
        \toks0=\e\e\e{\the\e\toks0\e&\e\multispan\e{\the\count1}\leaders\hrule\hfill}
        \fi \count1=0 }
    \def\putout##1{\ifx##1\end \putspace \putrule \let\next=\relax 
        \else \let\next=\putout
            \ifx##1- \advance\count1 by 2 \putspace
            \else    \advance\count0 by 2 \putrule \fi \fi \next}   }
\def\tablespec#1{
    \def\vdimens{\noexpand\tableheight}
    \def\tabby{\tabskip=0pt plus100pt minus100pt}
    \def\r{&################\tabby&\hfil################\unskip}
    \def\c{&################\tabby&\hfil################\unskip\hfil}
    \def\l{&################\tabby&################\unskip\hfil}
    \edef\templ{\noexpand\vdimens ########\unskip  #1 
         \unskip&########\tabskip=0pt&########\cr}
    \tabledelim
    \edef\body##1{ \vbox{
        \tabskip=0pt \offinterlineskip
        \halign to \twidth {\templ ##1}}} }

%%%%%%%%%%%%%%%%%%
\newbox\grsign \setbox\grsign=\hbox{$>$}
\newdimen\grdimen \grdimen=\ht\grsign
\newbox\laxbox \newbox\gaxbox
\setbox\gaxbox=\hbox{\raise.5ex\hbox{$>$}\llap
	{\lower.5ex\hbox{$\sim$}}}\ht1=\grdimen\dp1=0pt
\setbox\laxbox=\hbox{\raise.5ex\hbox{$<$}\llap
	{\lower.5ex\hbox{$\sim$}}}\ht2=\grdimen\dp2=0pt

\def\uJy{\ifmmode{\,\mu{\rm Jy}}\else$\,{\mu{\rm Jy}}$\fi}
\def\mJy{\ifmmode{\,{\rm mJy}}\else${\,{\rm mJy}}$\fi}
\def\MHz{\ifmmode{\,{\rm MHz}}\else{$\,{\rm MHz}$}\fi}
\def\GHz{\ifmmode{\,{\rm GHz}}\else{$\,{\rm GHz}$}\fi}
\def\solar{\ifmmode_{\mathord\odot}\else$_{\mathord\odot}$\fi}
\def\Msolar{\ifmmode{\, {\rm M\solar}}\else{${\, {\rm M\solar}}$}\fi}
\def\Rsolar{\ifmmode{\, {\rm R\solar}}\else{${\, {\rm R\solar}}$}\fi}
\def\kms{\ifmmode{\,{\rm km\,s^{-1}}}\else${\,{\rm km\,s^{-1}}}$\fi}
\def\kpc{\ifmmode{\,{\rm kpc}}\else${\,{\rm kpc}}$\fi}
\def\us{\ifmmode{\,\mu{\rm s}}\else$\,{\mu{\rm s}}$\fi}
\def\ms{\ifmmode{\,{\rm ms}}\else$\,{{\rm ms}}$\fi}
\def\y{\ifmmode{\,{\rm y}}\else$\,{\rm y}$\fi}
\def\h{\ifmmode{^{\rm h}}\else$^{\rm h}$\fi}
\def\m{\ifmmode{^{\rm m}}\else$^{\rm m}$\fi}
\def\s{\ifmmode{^{\rm s}}\else$^{\rm s}$\fi}
\def\Lmin{\ifmmode{L_{min}}\else{$L_{min}$}\fi}

\input psfig.sty
\magnification=\magstep1
\singlespace
%\mediumspace
%\doublespace
\font\eightrm=cmr8

\font\fourrm=cmr5
% for FAX version send with mag=1370 and doublespace
%\magnification=1370

%\doublespace
%------------------------------------------------------------------------
%Specific Definitions
%------------------------------------------------------------------------
\font\lgh=cmbx10 scaled \magstep2
\def\hb{\hfill\break}
%Figure numbers

%Table Numbers

%------------------------------------------------------------------------
% THE TEXT STARTS HERE:
%------------------------------------------------------------------------

\noindent{\hfill Version: \today}
\smallskip
\hrule
\bigskip
%\line{\lgh Submillimetre imaging of distant radio galaxies: \hb}
%\line{\lgh witnessing the formation of massive cluster ellipticals\hb}
\line{\lgh The formation of cluster elliptical galaxies as\hb}
\line{\lgh revealed by extensive star formation\hb}

\bigskip
\bigskip

\line{J.\ A.\ Stevens$^*$, R.\ J.\ Ivison$^*$, J.\ S.\ Dunlop$\dag$, 
   Ian Smail$\ddag$,
   W.\ J.\ Percival$\dag$, D. H.\ Hughes$\S$ \hb}

\line{H.\ J.\ A.\ R\"ottgering$\Vert$, W.\ J.\ M.\ van Breugel$\P$
\& M.\ Reuland$\P$ \hb} 

\bigskip

\line{$^*$
      Astronomy Technology Centre, Royal Observatory, Blackford Hill,
      Edinburgh \hfill}
\line{$\;\,$ EH9 3HJ, UK \hfill}

\line{$\dag$
      Institute for Astronomy, University of Edinburgh, Blackford Hill, \hfill}
\line{$\;\,$ Edinburgh EH9 3HJ \hfill}

\line{$\ddag$
      Institute for Computational Cosmology, University of Durham,  \hfill}
\line{$\;\,$South Road, Durham DH1 3LE \hfill}

\line{$\S$
      Instituto Nacional de Astrofisica, Optica y Electronica,
      Apartado Postal 51 y 216, 72000 \hfill}
\line{$\;\,$ Puebla, Mexico \hfill}

\line{$\Vert$
      Leiden Observatory, PO Box 9513, 2300 Leiden, The Netherlands \hfill}

\line{$\P$
      Institute of Geophysics and Planetary Physics, Lawrence Livermore
      National Laboratory, \hfill}
\line{$\;\,$ PO Box 808, Livermore, CA$\,$94459, USA \hfill}

\bigskip
\hrule
\bigskip
\bigskip

\noindent {\bf 
The most massive galaxies in the present-day Universe are found to lie in the
centres of rich clusters. They have old, coeval stellar populations suggesting
that the bulk of their stars must have formed at early epochs in spectacular
starbursts\refto{Ee97} -- luminous phenomena at submillimetre
wavelengths\refto{De94}.  The most popular model of galaxy formation predicts
that these galaxies form in proto-clusters at high-density peaks in the early
Universe\refto{Ka96}. Such peaks are signposted by massive high-redshift radio
galaxies\refto{Wt94}.  Here we report deep submillimetre mapping of seven
high-redshift radio galaxies and their environments.  These data confirm not
only the presence of spatially extended massive star-formation activity in the
radio galaxies themselves, but also in companion objects previously undetected
at any wavelength. The prevalence, orientation, and inferred masses of these
submillimetre companion galaxies suggest that we are witnessing the synchronous
formation of the most luminous elliptical galaxies found today at the centres
of rich galaxy clusters.}

\vfill
\eject

Whilst existing submillimetre studies of high-redshift radio galaxies
(hereafter HzRGs) have shown that their star-formation rates are large enough
to build a massive galaxy in $<1$~Gyr\refto{De94,HDR97,Ae01,Re03} they have
provided no information on the spatial extent of this emission or on the
star-formation activity in their environments.  We have therefore mapped a
sample of seven objects with redshifts ranging from 2.2 to 4.3 at a wavelength
of 850$\,\mu$m with the Submillimetre Common-User Bolometer Array
(SCUBA)\refto{He99} on the James Clerk Maxwell Telescope (JCMT). The targets
were chosen from those sources found to be submillimetre bright in the previous
SCUBA surveys of HzRGs\refto{Ae01,Re03}. Our new maps illustrate the
distribution of dust-reradiated emission in and around the HzRGs on scales from
5$''$ to 160$''$, or 30$\,$kpc to 1$\,$Mpc.  We illustrate the seven
submillimetre maps from this survey in Figure~1; the orientation of the radio
jets of each HzRG is represented by tick marks on these maps.

One of the most striking aspects of the submillimetre maps is that the dust
emission from the central radio galaxy is resolved in at least five of the
seven sources -- even with the coarse beam of the JCMT. In Figure~2 and Table~1
we present simple two-dimensional Gaussian fits to the data which, while not
giving a true reflection of the physical situation, at least provide a
quantitative measure of the spatial extent of the dust emission.  This emission
is sometimes in the form of several partially-resolved or merged clumps
(typified by 8C$\,$1909+722), sometimes in an apparently smoother distribution
(e.g.\ 4C$\,$60.07), and is more extended than the radio emission in most
cases. The extent of the dust emission ranges from 50 to 250~kpc, a physically
interesting size because (1) the corresponding half-light radii (30--150~kpc)
are equivalent to those of brightest cluster galaxies in the local
Universe\refto{Gea96}, and (2) gas-dynamical simulations of major galaxy mergers
predict that star formation should peak when the galaxies are separated on
approximately this scale.\refto{MH96} Higher resolution, high signal-to-noise
millimetre/submillimetre imaging observations are required to investigate these
possibilities.

A second striking feature of our SCUBA maps is that several of the fields also
contain serendipitous detections of new, luminous submillimetre galaxies.  In
Table~1 we list the properties of those companion galaxies with peak
signal-to-noise ratio $>$\ 4. Note that several of these companions are, like
the HzRGs, resolved at 850$\,\mu$m. The source density of companions is also
higher than found in `blank-field' surveys; current submillimetre number
counts\refto{SIBK02} indicate that we should find, on average, about one random
submillimetre source per SCUBA field at a flux level of $S_{850} >\ $5--6\ mJy
compared to our detection of approximately twice this number (Table~1).
Furthermore, the 4C\ 41.17 and 8C\ 1909+722 fields each contain a companion
source with $S_{850} > 20$\ mJy. By comparison, the biggest blank field survey
conducted to date\refto{Sea02}, which has an area $\sim7$ times that of our
own, contains no robust sources with S$_{850} > 15$\ mJy.

Arguably the most surprising feature of the maps is the observed aligments
between the HzRG radio axis and (1) the submillimetre emission (2) the
brightest submillimetre `companions'. The significance of these alignments is
presented and quantified in Figure~3. These results can be interpreted as
follows. First, while one might be tempted to conclude that the first alignment
effect is indicative of jet-induced star formation similar to the radio-UV
alignment effect previously reported in HzRGs\refto{CMvB87,Mc87}, the fact that
in all but two cases, the radio source is also aligned with submillimetre
source on scales well beyond the size of the radio emission indicates that this
is unlikely to be the complete explanation.  Moreover, this model clearly
cannot explain why the submillimetre emission from the HzRGs appears to be
aligned with positions of the brightest companion sources in their vicinity (in
at least 4 out of 7 cases).

There is one model that can explain {\it both\/} alignment effects.  We
propose that the brightest submillimetre companions trace the large-scale
structure around the HzRG and that these directions thus contain the densest
cross-sections of gas.  Next, by selecting some of the very brightest known
radio sources at this epoch we will then have (inadvertently, but probably
inevitably) selected sources which happen to have produced jets aligned with
the densest regions of gas, thereby producing very effective working surfaces
and the brightest hot spots. Such a selection effect has been suggested before
to explain apparent large-scale optical-radio alignment effects at lower
redshift.\refto{Es92,Wt94}

Of course, an important corollary to this interpretation is that the brightest
companion sources seen in the submillimetre maps must lie at the same redshift
as the radio galaxies, occupying the same large-scale structure.  However, the
fact that extension of this analysis to the second brightest apparent companion
object does not reveal a significant alignment effect shows that not all of the
objects seen in these images need lie at the same redshift as, and be
physically associated with the radio galaxies.  This is consistent with our
earlier estimate of the rate of contamination in these fields by unrelated
submillimetre sources. Thus the analysis of these possible alignment effects
indicates that the submillimetre emission from the radio galaxies themselves,
and that of the brightest companions, is tracing the large scale structure
around radio galaxies at $z \simeq 3$. We note that for 53W002, a HzRG not
included in our sample, it has been shown with optical spectroscopy that the
brightest submillimetre companion is indeed at the same redshift as the radio
galaxy.\refto{Se03} A similar conclusion can be inferred for PKS~$1138-262$
where the brightest submillimetre companion is coincident with one of 5 actuve
galactic nuclei (AGN) forming a radio-aligned filament in the plane of the sky,
and which has the same redshift as the HzRG based on narrow-band imaging of
redshifted Ly$\alpha$ emission.\refto{Pea02}

Since we have shown that these fields contain overdensities of submillimetre
sources, it is of interest to ask what is the typical mass of these galaxies.
We can do this by assuming that they lie at the same redshift as the HzRGs.
The dust masses of the HzRGs, calculated from their 850-$\mu$m
fluxes\refto{Hd83} are given in Table~1. Using standard assumptions we can
convert these dust masses into total gas masses, and hence estimate the masses
of their associated dark matter haloes (see table caption for details).  We
estimate that they reside in dark matter halos with masses in excess of
$10^{12}\, {\rm M_{\odot}}$.  Since the submillimetre flux densities of the
companions are similar to those of the HzRGs we can infer that they also reside
in dark matter halos with masses in excess of $10^{12}\, {\rm M_{\odot}}$.  We
therefore conclude that these regions can contain more than one very massive
and gas-rich galaxy of $>10^{12}\, {\rm M_{\odot}}$. Can we reconcile this
result with the hierarchical models of galaxy and structure formation?
 
By design we have selected highly-biased regions of the
high-redshift universe by imaging around some of the most luminous radio
galaxies at these epochs, which are expected to host some of the most massive
black holes. We can attempt to quantify the impact of this bias by
exploring the predictions of numerical simulations of the collapse of Cold Dark
Matter haloes within a $\Lambda$-dominated universe. The details and results of
these simulations are given in Figure~4.  These simulations do indeed appear to
be in accord with the observations, i.e. both data and theory suggest that the
submillimetre companions revealed by our SCUBA imaging are associated with
dark-matter halos more massive than $10^{12}\, {\rm M_{\odot}}$. This
conclusion is, however, subject to an assumption about the duty cycles of the
submillimetre luminous phase; if true it would imply that the vast majority of
such massive halos are actively forming stars at the epochs sampled by these
images. This is not unreasonable, especially since we have deliberately
targetted fields in which at least one massive object (i.e.\ the HzRG) is
actively engaged in intense star formation.

Pursuing this comparison one step further, we can explore the
predicted and observed distribution of companion objects in this high-mass
range. In fact, while the average number of such companions in the simulated
SCUBA images centred on the HzRG is 1, this average arises from a skewed
distribution. As shown in Figure~4, of the 40 high-mass haloes investigated,
50\% have no such massive companions, with the average of 1 resulting from the
fact that, of the remaining half, 40\% have 2 companions or more. We have
over-plotted in Figure~4 the corresponding histogram of companion incidence for
the 7 HzRGs imaged in this study (after statistically correcting for field
contamination). This comparison at least illustrates that the distribution of
companion incidence in our data is consistent with the prediction of the
simulations, i.e.\ typically we see either no companion, or $\geq 2$
companions.

The average number of companions, the distribution of companion incidence, and
the inferred baryonic gas masses of the companions are all most consistent with
the interpretation that the SCUBA sources uncovered in this study are the
progenitors of massive present-day cluster ellipticals. In this case, since
elliptical galaxies are known to contain massive black holes with mass
proportional to that of the spheroid,\refto{Ma98} it is reasonable to assume
that the black hole and stellar mass grow coevally from the same gas
reservoir. This in turn suggests that the companion objects should contain
buried AGN.\refto{KH00,Pea01} First results show a high rate of correspondence
between the submillimetre companions and luminous X-ray
sources,\refto{Se03,Sea03} suggesting that this is indeed the case. We are
presently pursuing high-resolution follow-up observations of these fields at
millimetre and optical/near-infrared wavelengths. The former will yield
information on source structure while the latter will reveal counterparts for
spectroscopic redshift determination on 10-m class telescopes.

%We conclude, therefore, that in these maps we are witnessing the final stages of
%the formation of the stellar populations of the most luminous elliptical
%galaxies found in present-day rich clusters.  Both the epoch of this activity
%(i.e.\ mostly at $z > 3$) and the fact that the main galaxies are forming stars
%at the same time, are consistent with the well-established age and coevality of
%brightest cluster members.\refto{Ee97}

\vfill\eject

%-----------------------------------------------------------------------
%REFERENCES
%-----------------------------------------------------------------------

\def\apj{Astrophys.\ J.}

\def\aj{Astron.\ J.}
\def\mnras{Mon.\ Not.\ R.\ Astron.\ Soc.}
\def\qjras{Qua.\ J.\ R.\ Astron.\ Soc.}
\def\aap{Astron. \& Astrophys.}

\bigskip

\noindent
{\bf References.}

%\refis{Ae93}
%       Arag\'on-Salamanca, A., Ellis, R.\ S., Couch, W.\ J.\ \&
%       Carter, D.
%       Evidence for systematic evolution in the properties of galaxies
%       in distant clusters
%       {\it \mnras}\ {\bf 262,} 764--794 (1993).

%\refis{ACG95}
%      Andreani, P., Casoli, F. \& Gerin, M.
%      CO, HI and cold dust in a sample of IRAS galaxies.
%      {\it \aap}\ {\bf 300,} 43--57 (1995).

\refis{Ae01}
       Archibald, E.\ N.\ et al.
       A submillimetre survey of the star-formation history of radio
       galaxies.
       {\it \mnras}\ {\bf 323,} 417--444 (2001).

%\refis{BW87}
%       Baron, E.\ \& White, S.\ D.\ M.
%       The appearance of primeval galaxies.
%       {\it Astrophys.\ J.}\ {\bf 322,} 585--596 (1987).

%\refis{Baugh98}
%       Baugh, C.\ M., Cole, S., Frenk, C.\ S.\ \& Lacey, C.\ G.
%       The epoch of galaxy formation.
%       {\it \apj}\ {\bf 498,} 504--521 (1998).

%\refis{Baugh99}
%       Baugh, C.\ M., Cole, S., Frenk, C.\ S., Benson, A.\ J.\ \& Lacey, C.\ G.
%       Early-type galaxies in the hierarchical universe. In: Star formation
%       in early-type galaxies, eds Carral, P.\ \& Cepa, J., ASP Conf.\ Series
%       163, p.\ 227 (1999).

%\refis{BI02}
%       Barvainis, R.\ E.\ \& Ivison, R.\ J.
%       A submillimeter survey of gravitationally lensed quasars.
%       {\it \apj}\ submitted.

%\refis{BLE92a}
%       Bower, R.\ G., Lucey, J.\ R.\ \& Ellis, R.\ S.
%       Precision photometry of early-type galaxies in the Coma and
%       Virgo clusters: A test of the universality of the
%       colour-magnitude relation. I.\ --- The data.
%       {\it \mnras}\ {\bf 254,} 589--613 (1992).

%\refis{BLE92b}
%       Bower, R.\ G., Lucey, J.\ R.\ \& Ellis, R.\ S.
%       Precision photometry of early-type galaxies in the Coma and
%       Virgo clusters: A test of the universality of the
%       colour-magnitude relation. I.\ --- Analysis.
%       {\it \mnras}\ {\bf 254,} 601--613 (1992).

\refis{dBr03}
      De Breuck, C.\ et al.
      CO emission and associated HI absorption from a massive gas reservoir
      surrounding the $z=3$ radio galaxy B3$\,$J2330+393.
      {\it \aap}\ {\bf 401,} 911--925 (2003).

%\refis{vBe99}
%       van Breugel, W.\ et al.
%       A radio galaxy at $z$ = 5.19.
%       {\it \apj}\ {\bf 518,} L61--L64 (1999).

%\refis{vBe98}
%       van Breugel, W.\ J.\ M.\ et al.
%       Morphological evolution in high redshift radio galaxies and the
%       formation of giant elliptical galaxies.
%       {\it \apj}\ {\bf 502,} 624--629 (1998).   

%\refis{BS01}
%       Burkert, A.\ \& Silk, J.
%       Star formation-regulated growth of black holes in protogalactic
%       spheroids.
%       {\it \apj}\ {\bf 554,} L151--L154 (2001).

%\refis{Ce98}
%       Carilli, C.\ L.\ et al.
%       An X-ray cluster at redshift 2.156?
%       {\it \apj}\ {\bf 494,} L143--L146 (1998).

\refis{CMvB87}
       Chambers, K.\ C., Miley, G.\ K.\ \& van Breugel, W.
       Alignment of radio and optical orientations in high-redshift
       radio galaxies.
       {\it Nature}\ {\bf 329,} 604--606 (1987).

%\refis{De97} Dey, A., van Breugel, W., Vacca, W.\ D. \& Antonucci, R.
%      Triggered star formation in a massive galaxy at $z=3.8$: 4C\ 41.17.
%      {\it \apj}\ {\bf 490,} 698--709 (1997).  

%\refis{vDF01}
%      van Dokkum, P. G.\ \& Franx, M.
%      Morphological evolution and the ages of early-type galaxies in clusters.
%      {\it \apj}\ {\bf 553,} 90--102 (2001).

%\refis{JSD}
%	Dunlop, J.\ S.
%	Sub-mm clues to elliptical galaxy formation.
%	In: `Deep millimeter surveys', eds Lowenthal, J.\ \& Hughes D.,
%        World Scientific, astro-ph/0011077 (2001).

%\refis{DL84}
%       Draine, B.\ T.\ \& Lee, H.\ M.
%       Optical properties of interstellar graphite and silicate grains.
%       {\it \apj}\ {\bf 285,} 89--108 (1984).

%\refis{DS98}
%      Downes, D.\ \&  Solomon, P.\ M.
%      Rotating nuclear rings and extreme starbursts in ultraluminous galaxies.
%      {\it \apj}\ {\bf 507,} 615--654 (1998).

%\refis{DY90}
%      Devereux, N.\ A.\ \& Young, J.\ S.
%      The gas/dust ratio in spiral galaxies.
%      {\it \apj}\ {\bf 359,} 42--56 (1990)

%\refis{DP93}
%      Dunlop, J.\ S.\ \& Peacock, J.\ A.
%      Luminosity dependence of optical activity and alignments in
%      radio galaxies.
%      {\it \mnras}\ {\bf 263,} 936--966 (1993).

\refis{De94}
      Dunlop, J.\ S.\ et al.
      Detection of a large mass of dust in a radio galaxy at redshift $z=3.8$.
      {\it Nature}\ {\bf 370,} 347--349 (1994).

\refis{Du03}
      Dunlop, J.\ S.\ et al.
      Quasars, their host galaxies, and their central black holes.
      {\it \mnras}\ {\bf 340,} 1095--1135 (2003).

\refis{Es92}
      Eales, S.\ A.
      A new theory for the alignment effect.
      {\it \apj}\ {\bf 397,} 49--54 (1992).

\refis{Ee97}
       Ellis, R.\ S.\ et al.
       The homogeneity of spheroidal populations in distant clusters.
       {\it \apj}\ {\bf 483,} 582--596 (1997).

%\refis{DW74}
%       Davis, M.\ \& Wilkinson, D.\ T.
%       Search for primeval galaxies.
%       {\it Astrophys.\ J.}\ {\bf 192,} 251--260 (1974).

%\refis{F64}
%       Field, G.\ B.
%       Quasi-stellar radio sources as spherical galaxies in the process
%       of formation.
%       {\it Astrophys.\ J.}\ {\bf 140,} 1434--1444 (1964).

%\refis{Ge00}
%       Gebhardt, K.\ et al.
%       A relationship between nuclear black hole mass and galaxy
%       velocity dispersion
%       {\it \apj}\ {\bf 539,} L13--L16 (2000).

%\refis{Ge98}
%      Genzel, R.\ et al.
%      What powers ultraluminous IRAS galaxies?
%      {\it \apj}\ {\bf 498,} 579--605 (1998). 

\refis{Go75}
      Gorenstein, P.
      Empirical relation between interstellar X-ray absorption and optical
      extinction.
      {\it \apj}\ {\bf 198,} 95--101 (1975).

%\refis{GSM96}
%       Giavalisco, M., Steidel, C.\ C.\ \& Macchetto F.\ D.
%       Hubble Space Telescope imaging of star-forming galaxies at redshifts
%       $z>3$.
%       {\it Astrophys.\ J.}\ {\bf 470,} 189--194 (1996).

\refis{Gea96}
      Graham, A., Lauer T.\ R., Colless M., \& Postman M.
      Brightest cluster galaxy profile shapes.
      {\it \apj}\ {\bf 465,} 534--547 (1996).

%\refis{He90}
%       Heckman, T.\ M., Armus, L.\ \& Miley, G.\ K.
%       On the nature and implications of starburst-driven galactic superwinds.
%       {\it Astrophys.\ J.\ Suppl.}\ {\bf 74,} 833--868 (1990).

\refis{Hd83}
       Hildebrand, R.\ H.
       The determination of cloud masses and dust characteristics from
       submillimetre thermal emission.
       {\it \qjras}\ {\bf 24,} 267--282 (1983).

\refis{He99}
       Holland, W.\ S.\ et al.
       SCUBA: a common-user submillimetre camera operating on the
       James Clerk Maxwell Telescope.
       {\it \mnras}\ {\bf 303,} 659--672 (1999).

%\refis{He98}
%       Hughes, D.\ H.\ et al.
%       High-redshift star formation in the Hubble Deep Field revealed
%       by a submillimetre-wavelength survey.
%       {\it Nature}\ {\bf 394,} 241--247 (1998).

\refis{HDR97}
       Hughes, D.\ H., Dunlop, J.\ S.\ \& Rawlings, S.
       High-redshift radio galaxies and quasars at submillimetre
       wavelengths: assessing their evolutionary status.
       {\it \mnras}\ {\bf 289,} 766--782 (1997).

\refis{Ie00}
       Ivison, R.\ J.\ et al.
       An excess of submillimeter sources near 4C$\,$41.17: a candidate
       proto-cluster at $z$ = 3.8?
       {\it Astrophys.\ J.}\ {\bf 542,} 27--34 (2000).

%\refis{Ie95}
%      Ivison, R.\ J.\
%      Detection of dust in the most distant known radio galaxy.
%      {\it \mnras}\ {\bf 275,} L33--L36 (1995).

%\refis{Ie01}
%       Ivison, R.\ J., Smail, I., Frayer, D.\ T., Kneib, J.-P.\ \&
%       Blain, A.\ W.
%       Locating the starburst in the SCUBA galaxy, SMM$\,$J14011+0252.
%       {\it \apj}\ {\bf 461,} L45--L49 (2001).

%\refis{Je01}
%      Jarvis, M.\ J.\ et al.
%      A sample of 6C radio sources designed to find objects at $z>4$
%      -III. Imaging and the radio galaxy $K-z$ relation.
%      {\it \mnras}\ {\bf 326,} 1563--1584 (2001). 

%\refis{Ke98}
%       Kneib J.-P.\ et al.
%       Modelling the Cloverleaf: contribution of a galaxy cluster at
%       $z\sim$ 1.7.
%       {\it \aap}\ {\bf 329,} 827--839 (1998).

%\refis{Ke00}
%       Kurk, J.\ D.\ et al.
%       A search for clusters at high redshift --- I.\ Candidate
%       Ly$\,\alpha$ emitters near 1138$-$262 at $z$ = 2.2.
%       {\it \aap}\ {\bf 358,} L1--L4 (2000).

\refis{Ka96}
        Kauffmann, G.
        The age of elliptical galaxies and bulges in a merger model.
        {\it \mnras}\ {\bf 281,} 487--492 (1996).

\refis{KH00}
        Kauffmann, G. \& Haehnelt, M.
        A unified model for the evolution of galaxies and quasars.
        {\it \mnras}\ {\bf 311,} 576--588 (2000).
 
%\refis{KR95}
%        Kormendy, J. \& Richstone, D.
%        Inward bound -- The search for supermassive black holes in galactic
%        nuclei.
%        {\it \araa}\ {\bf 33,} 581--624 (1995).

%\refis{L69}
%       Larson, R.\ B.
%       A model for the formation of a spherical galaxy.
%       {\it \mnras}\ {\bf 145,} 405--422 (1969).

%\refis{L74}
%       Larson, R.\ B.
%       Dynamical models for the formation and evolution of spherical galaxies.
%       {\it \mnras}\ {\bf 166,} 585--616 (1974).

%\refis{Le98}
%       Lewis, G.\ F., Chapman, S.\ C., Ibata, R.\ A., Irwin, M.\ J.\ \&
%       Totten, E.\ J.
%       Submillimeter observations of the ultraluminous broad absorption line
%       quasar APM$\,$08279+5255.
%       {\it Astrophys.\ J.}\ {\bf 505,} L1--L5 (1998).

%\refis{Lutz}
%	Lutz, D.\ et al.
% 	The extended counterpart of submm source Lockman$\,$850.1.
%	{\it \aap}\ in press, astro-ph/0108131 (2001).

\refis{Ma98}
        Magorrian, J.\ et al.
        The demography of massive dark objects in galactic centers. 
        {\it \aj}\ {\bf 115,} 2285--2305 (1998).

%\refis{M76}
%       Meier, D.\ L.
%       The optical appearance of model primeval galaxies.
%       {\it Astrophys.\ J.}\ {\bf 207,} 343--350 (1976).

%\refis{McCe87a}
%       McCarthy, P.\ J.\ et al.
%       Extended Lyman-$\alpha$ emission in 3C$\,$326.1 --- a 100 kiloparsec
%       cloud of ionized gas at a redshift of 1.82.
%       {\it Astrophys.\ J.}\ {\bf 319,} L39--L44 (1987).

\refis{Mc87}
       McCarthy, P.\ J., van Breugel, W., Spinrad, H.\ \& Djorgovski, S.
       A correlation between the radio and optical morphologies of
       distant 3CR radio galaxies.
       {\it Astrophys.\ J.}\ {\bf 321,} L29--L33 (1987).

\refis{MH96}
       Mihos, J.\ C., \& Hernquist, L.
       Gasdynamics and starbursts in major mergers.
       {\it \apj}\ {\bf 464,} 641--663 (1996).

%\refis{McLe99}
%       McLure, R.\ J.\ et al.
%       A comparative {\it HST} imaging study of the host galaxies of
%       radio-quiet quasars, radio-loud quasars and radio galaxies --- I
%       {\it \mnras}\ {\bf 308,} 377--404 (1999).

%\refis{McMe94}
%       McMahon, R.\ G., Omont, A., Bergeron, J., Kreysa, E.\ \&
%       Haslam, C.\ G.\ T.
%       1.25-mm continuum observations of very high-redshift QSOs: Is
%       there dust at $z$ = 4.69?
%       {\it \mnras}\ {\bf 267,} L9--L12 (1994).

%\refis{Oe96}
%       Ohta, K.\ et al.
%       Detection of molecular gas in the quasar BR$\,$1202$-$0725 at
%       redshift $z$ = 4.69.
%       {\it Nature}\ {\bf 382,} 426--428 (1996).

\refis{Om96}
       Omont, A.\ et al.
       Molecular gas and dust around a radio-quiet quasar at redshift
       4.69.
       {\it Nature}\ {\bf 382,} 428--431 (1996).

\refis{Pea01}
       Page, M.\ J., Stevens, J.\ A., Mittaz, J.\ P.\ D.\ \& Carrera,
       F.\ J.
       Submillimetre evidence for the coeval growth of massive black
       holes and galaxy bulges.
       {\it Science}\ {\bf 294,} 2516--2518 (2001).

\refis{Pe00}
       Papadopoulos, P.\ P.\ et al.
       CO(4--3) and dust emission in two powerful high-$z$ radio galaxies,
       and CO lines at high redshifts.
       {\it Astrophys.\ J.}\ {\bf 528,} 626--636 (2000).

%\refis{PICL01}
%       Papadopoulos, P., Ivison, R., Carilli, C.\ \& Lewis, G.
%       A massive reservoir of low-excitation molecular gas at high redshift.
%       {\it Nature}\ {\bf 409,} 58--60 (2001).

%\refis{PP67}
%       Partridge, R.\ B.\ \& Peebles, P.\ J.\ E.
%       Are young galaxies visible?
%       {\it Astrophys.\ J.}\ {\bf 147,} 868--886 (1967).

%\refis{Pent01}
%       Pentericci, L.\ et al.
%       NICMOS observations of high-redshift radio galaxies: witnessing the
%       formation of bright elliptical galaxies?
%       {\it Astrophys.\ J.\ Suppl.}\ {\bf 135,} 63--85 (2001).

%\refis{Pent00}
%       Pentericci, L.\ et al.
%       A search for clusters at high redshift --- II.\ A proto cluster
%       around a radio galaxy at $z$ = 2.16.
%       {\it \aap}\ {\bf 361,} L25--L28 (2000).

\refis{Pea02}
       Pentericci, L.\ et al.
       A Chandra study of X-ray sources in the field of the z=2.16 radio galaxy
       MRC 1138-262
       {\it \aap}\ {\bf 396,} 109--115 (2002). 

\refis{Pe03}
      Percival, W.\ J., Scott, D., Peacock, J.\ A.\ \& Dunlop, J.\ S.
      The clustering of halo mergers.
      {\it \mnras}\ {\bf 338,} L31--L35 (2003). 

%\refis{Rich98}
%	Richstone, D.\ et al.
%	Supermassive black holes and the evolution of galaxies.
%	{\it Nature}, {\bf 395,} 14--19 (1998).

%\refis{Re95}
%      Renzini, A.
%      Stellar dating and formation of galactic spheroids.
%      IN: `Stellar populations' Eds van der Kruit. P.\ C., Gilmore,\ G.,
%      IAUS no. {\bf 164,} p. 325 (1995).

\refis{Re03}
      Reuland, M., R\"{o}ttgering, H. \& van Breugel, W.
      SCUBA observations of high redshift radio galaxies
      IN: `Radio Galaxies: Past, present and future.'  
      Elsevier Science, in press (astro-ph/0303321)

\refis{Re03b}
      Reuland, M.\ et al.
      An obscured radio galaxy at high redshift.
      {\it \apj}\ {\bf 582,} L71--L74 (2003).

%\refis{RR00}
%       Rowan-Robinson, M.
%       Hyperluminous infrared galaxies.
%       {\it \mnras}\ {\bf 316,} 885--900 (2000).

%\refis{Se88}
%       Sanders, D.\ B.\ et al.
%       Ultraluminous infrared galaxies and the origin of quasars.
%       {\it \apj}\ {\bf 325,} 74--91 (1988). 

\refis{Sea02}
      Scott, S.\ E.\ et al.
      The SCUBA 8-mJy survey - I. Submillimetre maps, sources and number counts.
      {\it \mnras}\ {\bf 331,} 817--838 (2002).

\refis{Se03}
       Smail, I.\ et al.
       A SCUBA galaxy in the protocluster around 52W002 at $z$=2.4.
       {\it \apj}\ {\bf 583,} 551--558 (2003).

\refis{Sea03}
      Smail, I.\ et al.
      {\it Chandra\/} detections of SCUBA galaxies around High-z radio sources.
      {\apj}\ submitted (2003).

\refis{SIBK02}
       Smail, I., Ivison, R.\ J.\ Blain, A.\ W. \& Kneib, J.-P.
       The nature of faint submillimetre-selected galaxies.
       {\it \mnras}\ {\bf 331,} 495--520 (2002).

%\refis{SM96}
%       Sanders, D.\ B.\ \& Mirabel, I.\ F.
%       Luminous infrared galaxies.
%       {\it \araa}\ {\bf 34,} 749--792 (1996).

%\refis{SR98}
%       Silk, J.\ \& Rees, M.\ J.
%       Quasars and galaxy formation.
%       {\it \aap}\ {\bf 331,} L1--L4 (1998).

%\refis{SIB}
%       Smail, I., Ivison, R.\ J.\ \& Blain, A.\ W.
%       A deep sub-millimeter survey of lensing clusters: a
%       new window on galaxy formation and evolution.
%       {\it Astrophys.\ J.}\ {\bf 490,} L5--L8 (1997).

%\refis{SED98}
%       Stanford, S.\ A., Eisenhardt, P.\ R.\ \& Dickinson, M.
%       The evolution of early-type galaxies in distant clusters.
%       {\it \apj}\ {\bf 492,} 461--479 (1998).

%\refis{Steidel}
%	Steidel, C.\ C., Giavalisco, M., Pettini, M., Dickinson, M.,
%	Adelberger, K.\ L.
%	Spectroscopic confirmation of a population of normal star-forming
%	galaxies at redshift $z>3$
%	{\it \apj}\ {\bf 462,} L17--L20 (1996).

\refis{Wt94}
      West M.\ J.
      Anisotropic mergers at high redshifts: the formation of cD galaxies and
      powerful radio sources.
      {\it \mnras} {\bf 268,} 79--102 (1994).

%\refis{WF91}
%        White, S.\ D.\ M.\ \& Frenk, C.\ S.
%        Galaxy formation through hierarchical clustering.
%        {\it \apj}\ {\bf 379,} 52--79 (1991).

%\refis{WR78}
%       White, S.\ D.\ M.\ \& Rees, M.\ J.
%       Core condensation in heavy halos -- a two-stage theory for
%       galaxy formation and clustering.
%       {\it \mnras}\ {\bf 183,} 341--358 (1978).

\endreferences

%\vfill\eject

\bigskip
\noindent
{\bf Acknowledgements.}

\noindent The JCMT is run by the Joint Astronomy Centre on behalf of
the Particle Physics \& Astronomy Research Council (PPARC). JAS, JSD, IRS and
WJP would like to acknowledge funding from PPARC, the Royal Society and the
Leverhulme Trust. The work of WvB and MR was performed under the auspices of
the U.S. Department of Energy, National Nuclear Security Administration by the
University of California, Lawrence Livermore National Laboratory.

\bigskip

\noindent
Correspondence and requests for materials should be addressed to Jason Stevens
(jas@roe.ac.uk).

\vfill\eject
\smallskip

\noindent{\bf Tables.}

{\fourrm \noindent {\bf Table 1. Target fields.}

\halign{
\quad # \hfil & \quad # \hfil & \quad # \hfil & \quad # \hfil & \quad # \hfil & \quad # \hfil &
\quad # \hfil & \quad # \hfil  & \quad # \hfil \cr
Target &Coordinates$^a$ &(J2000) &$z$&$S_{\rm peak}^b$&$S_{\rm total}^b$ &M$_{\rm d}^c$&FWHM/P.A.$^d$&Radio size/P.A.$^e$ \cr 
name  &R.A. &  Dec.   & &(mJy)&(mJy)& (M$_{\odot}$) &($''$/$^{\circ}$)&($''$/$^{\circ}$)\cr
\noalign{\medskip}
8C$\,$1909+722  &19 08 23.30 &+72 20 10.4&3.54
&20.5$\pm$1.2&34.9$\pm$3.0&1.2$\times$10$^9$&26$\times$9@18 &14@15\cr
1. & 19 08 27.47 & +72 19 28.0 & $\ldots$ & 17.6$\pm$1.2 & 23.0$\pm$2.5 & 
$\ldots$ & 12$\times$8@169& $\ldots$ \cr
2. & 19 08 29.31 & +72 20 49.6 & $\ldots$ & 6.5$\pm$1.2 & 8.7$\pm$2.4 &
$\ldots$ & $\ldots$ & $\ldots$ \cr
3. & 19 08 16.12 & +72 20 24.0 & $\ldots$ & 6.5$\pm$1.2 & 4.3$\pm$2.1 &
$\ldots$ & $\ldots$ & $\ldots$ \cr

8C$\,$1435+635  &14 36 37.33 &+63 19
13.1&4.261&5.9$\pm$1.0&6.0$\pm$2.1&3.3$\times$10$^8$&7$\times$0@159 &4@156\cr
1. & 14 36 32.46 & +63 20 02.5 & $\ldots$ & 4.3$\pm$1.0 & 3.9$\pm$1.8 &
$\ldots$ & $\ldots$ & $\ldots$ \cr

B3$\,$J2330+3927  &23 30 24.91 &+39 27
11.2&3.086&15.6$\pm$1.1&22.2$\pm$2.7&1.0$\times$10$^9$&16$\times$7@130 &2@149\cr
1. & 23 30 19.14 & +39 27 03.0 & $\ldots$ & 6.9$\pm$1.1 & 8.2$\pm$1.9 & 
$\ldots$ & $\ldots$ & $\ldots$ \cr
2. & 23 30 20.52 & +39 26 57.7 & $\ldots$ & 5.0$\pm$1.1 & 3.5$\pm$1.4 &
$\ldots$ & $\ldots$ & $\ldots$ \cr

PKS$\,$1138-262    &11 40 48.25 &-26 29
10.1&2.156&5.9$\pm$1.1&6.7$\pm$2.4&4.6$\times$10$^8$&30$\times$0@72 &11@86\cr 
1. & 11 40 53.38 & -26 29 11.9 & $\sim$2.16 & 9.9$\pm$1.1 & 7.8$\pm$2.2 &
$\ldots$ & 17$\times$0@60& $\ldots$ \cr
2. & 11 40 45.80 & -26 29 56.6 & $\ldots$ & 5.9$\pm$1.1 & 3.1$\pm$1.6 & 
$\ldots$ & $\ldots$ & $\ldots$ \cr
3. & 11 40 45.61 & -26 29 06.6 & $\ldots$ & 4.6$\pm$1.1 & 2.2$\pm$1.4 &
$\ldots$ & $\ldots$ & $\ldots$ \cr

4C$\,$60.07     &05 12 54.80 &+60 30
51.7&3.788&21.6$\pm$1.3&23.8$\pm$3.5&1.3$\times$10$^9$&11$\times$4@110&16@107\cr 
1. & 05 12 46.52 & +60 30 35.8 & $\ldots$ & 5.9$\pm$1.3 & 6.3$\pm$2.1 &
$\ldots$ & $\ldots$ & $\ldots$ \cr 

4C$\,$41.17     &06 50 52.15 &+41 30
30.8&3.792&12.3$\pm$1.2&12.0$\pm$2.3&7.2$\times$10$^8$&9$\times$6@97 &20@48\cr
1. & 06 50 51.52 & +41 30 01.5 & $\ldots$ & 12.2$\pm$1.2 & 21.2$\pm$2.9 &
$\ldots$ & 17$\times$15@54 & $\ldots$ \cr 
2. & 06 50 49.25 & +41 30 01.5 & $\ldots$ & 7.1$\pm$1.2 & 6.2$\pm$1.9 & 
$\ldots$ & 8$\times$0@50 & $\ldots$ \cr

WN$\,$J0305+3525  &03 05 47.42 &+35 25
13.4&3$\pm$1&12.2$\pm$1.0&17.6$\pm$2.4&6.9$\times$10$^8$&19$\times$9@9&2@64\cr 
}}

{\eightrm

\noindent
$^a$ Coordinates of the HzRG companion sources are measured from the
submillimetre maps. They are accurate to 4$-$5 arcseconds.

\noindent
$^b$ $S_{\rm peak}$ is the flux density per beam calculated at the peak of the
dust emission.  $S_{\rm total}$ is the flux density of total dust emission
calculated in an aperture. A calibration uncertainty of $\sim10$\% is not
included in these values. For some sources, the higher total flux densities
reflect the extended nature of the dust emission, consistent with the finding
that interferometric millimetre flux densities are often smaller than those
calculated from single dish measurements.\refto{Pe00,dBr03}

\noindent
$^c$ Dust mass calculated from the peak fluxes, and assuming a dust emissivity
index, $\beta$=2.0; a dust temperature, T$_{\rm d}$=40 K typical of HzRGs; a
mass absorption coefficient, k$_{\rm d}$(850$\,\mu$m)=0.076 m$^2\,$kg$^{-1}$;
and an $\Omega_{\rm m}=$0.3, $\Omega_\Lambda=$0.7, $h=$0.7 cosmology. Our
selection criteria result in a small spread in the calculated dust masses; an
average value is $\sim8\times10^8$\ M$_{\odot}$.  We can convert this dust mass
into a conservative gas mass (molecular and atomic) assuming a standard
Galactic gas-to-dust mass ratio of 200\refto{Go75} giving $1.6\times10^{11}$\
M$_{\odot}$.  We can confirm this estimate using the measured CO luminosities
of three of the HzRGs, which can then be turned into molecular gas, and then
total gas, mass using the standard Galactic conversion factor between CO
luminosity and H$_2$ mass and a ratio of atomic-to-molecular gas of $\sim2$. In
this manner, gas masses of $1.3-2.2\times10^{11}$\ M$_{\odot}$ have been
estimated for 4C$\,$60.07, 8C$\,$1909+722 and
B3$\,$J2330+3927.\refto{Pe00,dBr03} Hence the gas masses from the two methods
are in reasonable agreement, and a representative estimate for the total gas
mass in a HzRG with a submillimetre flux density of 10--15 mJy is thus $\simeq
2 \times 10^{11}$\ M$_{\odot}$.  While this number is arguably uncertain by a
factor of a few, it is very hard to argue that it can be inflated by an order
of magnitude.\refto{HDR97} Assuming an order of magnitude more mass is in the
form of dark matter we estimate the halo masses to be $>10^{12}$\ M$_{\odot}$.

\noindent
$^d$ From a 2-dimensional Gaussian fit to the 850$\,\mu$m emission, deconvolved
from the beam. All P.A.s are measured east of north. For the companion galaxies
we give sizes for the brightest (highest S/N) sources only. A quoted size of
zero arcseconds means that the source was not resolved in that direction.

\noindent
$^e$ Quoted radio sizes are hotspot separations measured at 5\ GHz.

\noindent
$^f$ WNJ$\,$0305+35 lacks a robust spectroscopic redshift, but is thought to
lie at $z=3\pm1$.\refto{Re03b}

\vfill\eject

\noindent{\bf Figure Captions.}

{\eightrm \noindent {\bf Figure 1.} Continuum emission from dust in and around
seven high-redshift radio galaxies. The data for 4C$\,$41.17 have been
published previously.\refto{Ie00} Data were obtained using the
SCUBA\refto{He99} submillimetre camera on the James Clerk Maxwell Telescope
(JCMT) in its dual-wavelength mapping mode during 1998--2001 in the top 20
percentile of Mauna Kea weather conditions. The beam, after smoothing to a FWHM
of 14.4$''$ at 850$\,\mu$m, is shown bottom right in the form of a map of the
blazar, 3C~345, with contours at $-$40, $-$30, $-$10, +10, +30, +50, +70, +90\%
of the central peak.  The JCMT secondary mirror was chopped and nodded by
30$''$ in Right Ascension (R.A.), resulting in the $-$0.5/+1.0/$-$0.5 beam
profile. After all overheads, 30--40$\,$ks were spent integrating on each of
the fields, a total of $\sim$100$\,$hr before overheads, split down into
1280-\sec\ integrations separated by checks on focus, pointing accuracy and
atmospheric opacity.  All images have been cleaned with the smoothed
beam.\refto{Ie00} Submillimetre contours are shown at $-$3, 3, 4, 5, 6, 8, 10
$\times \sigma$, where $\sigma$ includes the contribution due to sources below
the detection threshold and to the chopping and nodding procedure and ranges
from 1 to 2$\,$mJy$\,$beam$^{-1}$. The galaxy with the most compact appearance,
4C$\,$41.17, actually subtends $\sim$7.5$''$ FWHM when deconvolved from the
beam. The tick marks around the central radio galaxies show the direction of
the kiloparsec scale radio jets.}

{\eightrm \noindent {\bf Figure 2.}  Submillimetre extent of the radio
galaxies.  Radial profiles of the dust emission compared with that of the beam
(solid line). The x-axis shows the radius from the centre of the radio galaxy
dust emission (arcseconds) and the y-axis shows the normalised flux density. An
accurate beam profile or point spread function (PSF) was measured with a
$\sim1$~hr mapping observation of the blazar 3C$\,$345 which had an 850$\mu$m
flux density of $>2$~Jy at the time. The image (shown in Figure~1) has a full
width half maximum (FWHM) of 14.4$''$.  All subsequent size measurements are
deconvolved from the PSF (see Table~1).  The reality of the extended nature of
the dust emission is established from observations of the quasar,
BR$\,$1202$-$0725, using the same technique and integration times used for the
radio galaxies to mimic any systematic errors.  BR$\,$1202$-$0725 serves as a
particularly conservative comparison object --- its dust continuum was
resolved\refto{Om96} by the IRAM interferometer into two components separated
by 4$''$ (P.A. 120$\pm$10$^{\circ}$). As expected, it appears relatively
compact (6.5$''$ $\times$ 5.5$''$ at P.A.\ 6$^{\circ}$) in our submillimetre
maps, and its radial profile is much more similar to that of the beam than to
those of the HzRGs. We conclude that our observing procedure does not produce
spurious morphologies, that the extended nature of the submillimetre emission
from the radio galaxies is real and occurs on scales that are typically
$\gg$4$''$.}

{\eightrm \noindent {\bf Figure 3.} Observed alignment effects.  The histograms
show the offset in degrees between the kiloparsec scale radio structure of the
HzRG and a) the position angle of the submillimetre source, b) the position
angle defined by the projected vector joining the radio source to the
brightest, spatially distinct submillimetre source, and c) the position angle
defined by the projected vector joining the radio source to the second
brightest, spatially distinct submillimetre source. In four cases these
submillimetre sources are the companions listed in Table~1. In the remaining
three cases one or both of the companions has a signal-to-noise ratio of 2 to
4; while these sources are not listed in Table~1 we use them in the analysis
for completeness.  We have applied the Kolmogorov Smirnov test to assess
whether the apparent alignment effect seen in the first two of these plots is
indeed statistically significant. Despite the small number statistics, we find
that both of these apparent alignment effects are significant at the 2-$\sigma$
level (as compared to a random distribution; $p=0.04$ and $p=0.05$
respectively) whereas the corresponding result for the third distribution is
not significant ($p=0.3$).}

{\eightrm \noindent {\bf Figure 4.} Comparison with hierarchical models.  The
predicted distribution of number of companion sources per SCUBA field (after
subtracting a background of 1 source per map for field contamination) from our
$\Lambda$--CDM simulations shown against the observed distribution (shaded
region). For the simulations we assume that the radio emission from the HzRG is
powered by black holes with mass $\simeq 10^9\, {\rm M_{\odot}}$\refto{Du03},
and therefore these lie at the heart of dark matter haloes with masses of
$\simeq 10^{13}\,{\rm M_{\odot}}$\refto{Ma98}. Allowing for intrinsic
scatter and uncertainties in this estimate, we therefore searched for
virialized dark matter haloes with masses in the range $12.5 < \log_{10}(M_{\rm
halo})/ {\rm M_{\odot}} < 13.5$ at $z \simeq 3$ within an N-body simulation of
a region of a $\Lambda$--CDM universe, covering a comoving volume of
($100h^{-1}$Mpc)$^3$. Parameters of the simulation were matched to the
``concordance model'' with $\Omega_{\rm m}=0.3$, $\Omega_\Lambda=0.7$, $h=0.7$;
further details of the simulation are given elsewhere
($\Lambda$CDM$_{100a}$)\refto{Pe03}. Within this simulation we identified 40
haloes which could be the counterparts of the HzRGs, and then investigated the
mass distribution of other virialized dark-matter haloes within a sphere of
projected diameter 3-arcmin (approximately the SCUBA field of view) around each
of these.  Within these regions we found typically 10 haloes with $M >
10^{11}\,{\rm M_{\odot}}$, but on average only 1 halo with $M > 10^{12}\,{\rm
M_{\odot}}$.}

\vfill
\eject

%{\bf Stevens-fig1}

\smallskip

\centerline{\hbox{\psfig{file=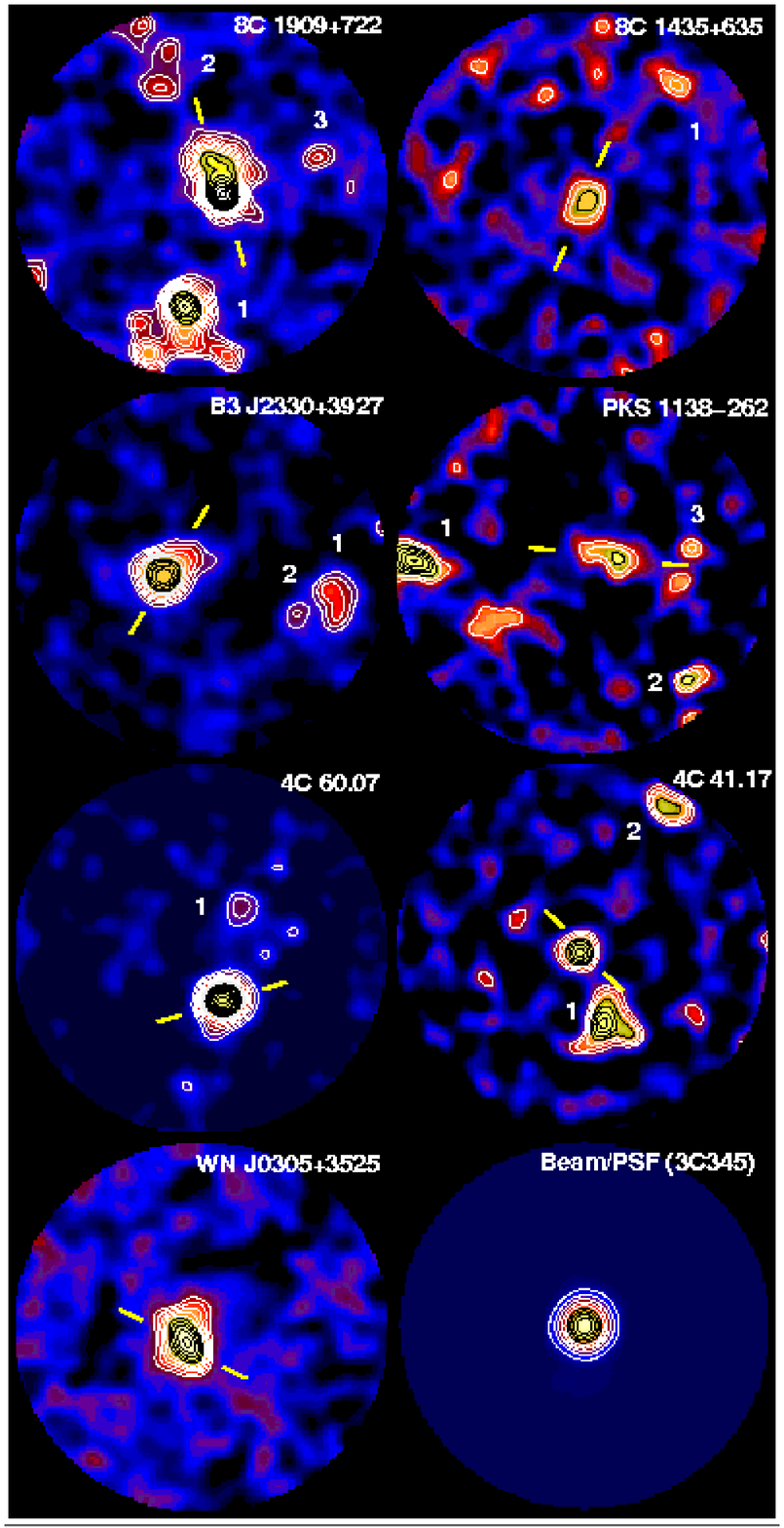,width=95mm,angle=0}}}

\vfill\eject

%{\bf Stevens-fig2}

\centerline{\hbox{\psfig{file=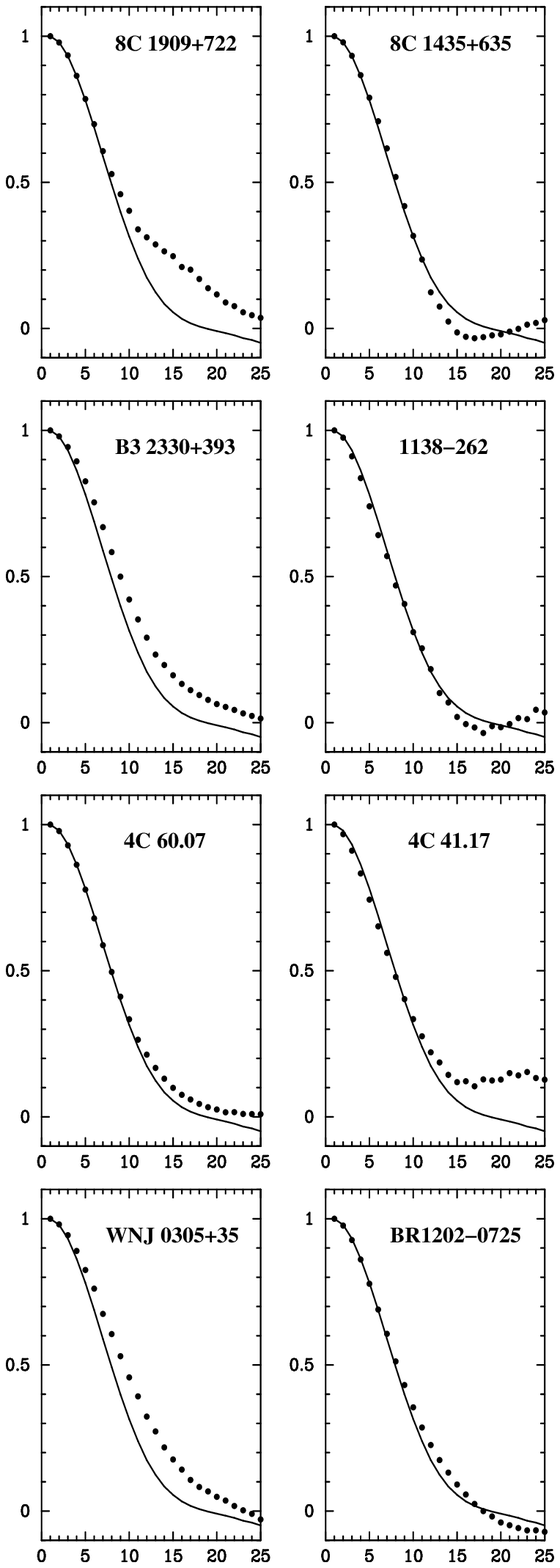,width=150mm,angle=0}}}

\vfill\eject

%{\bf Stevens-fig3}

\centerline{\hbox{\psfig{file=figure3.ps,width=70mm,angle=-90}}}

%\vfill\eject

\bigskip
\bigskip

%{\bf Stevens-fig4}

\centerline{\hbox{\psfig{file=figure4.ps,width=70mm,angle=0}}}

\vfill\bye